\documentclass{article}
\usepackage{algorithm,spconf,amsmath,amssymb,graphicx}
\usepackage{algpseudocode}

\title{\fontsize{11}{12} \selectfont Joint Transaction Transmission and Channel Selection in Cognitive Radio Based Blockchain Networks: A Deep Reinforcement Learning Approach}
\name{\fontsize{11}{12} \selectfont  Nguyen Cong Luong$^{1}$, Tran The Anh$^{1}$, Huynh Thi Thanh Binh$^{2}$, Dusit Niyato$^{1}$, Dong In Kim$^{3}$, and Ying-Chang Liang$^{4}$}
\address{\normalsize
 $^{1}$School of Computer Science and Engineering, Nanyang Technological University, Singapore\\
\normalsize  $^{2}$School of Information and Communication Technology, Hanoi University of Science and Technology, Vietnam\\
\normalsize  $^{3}$School of Information and Communication Engineering, Sungkyunkwan University, Korea\\
\normalsize $^{4}$Center for Intelligent Networking and Communications, University of Electronic Science and Technology of China, China\\
 }
\begin{document}
%
\maketitle
\begin{abstract}
To ensure that the data aggregation, data storage, and data processing are all performed in a decentralized but trusted manner, we propose to use the blockchain with the mining pool to support IoT services based on cognitive radio networks. As such, the secondary user can send its sensing data, i.e., transactions, to the mining pools. After being verified by miners, the transactions are added to the blocks. However, under the dynamics of the primary channel and  the uncertainty of the mempool state of the mining pool, it is challenging for the secondary user to determine an optimal transaction transmission policy. In this paper, we propose to use the deep reinforcement learning algorithm to derive an
optimal transaction transmission policy for the secondary user. Specifically, we adopt a Double Deep-Q Network (DDQN) that allows the secondary user to
learn the optimal policy.~The simulation results clearly show that
the proposed deep reinforcement learning algorithm outperforms
the conventional Q-learning scheme in terms of reward and learning speed.
\end{abstract}
\begin{keywords}
Cognitive radio, blockchain, IoT, channel access, deep reinforcement learning
\end{keywords}
\section{Introduction}
\label{sec:intro}

Cognitive radio has been adopted to support IoT data transmission from IoT devices to a centralized server or the cloud~\cite{khan2017cognitive}.~Specifically, the IoT devices act as the Secondary Users (SUs) accessing idle spectrum of the Primary Users (PUs), improving the IoT performance and enhancing PUs' spectrum utilization. Generally, the IoT data to support IoT services and applications involves multiple stakeholders including devices owners, service providers, and users. Hence, the traditional approach of maintaining the IoT data by a single entity, e.g., an IoT provider, has shown many limitations. Firstly, it lacks transparency and traceability, i.e., the data can be modified arbitrarily by unknown persons and applications. Secondly, security is limited as it has to rely on a single entity which can be an easy target of cyber attacks. Thirdly, efficiency, speed, and reliability are low because of a bottleneck and a single point of failure. This calls for a novel solution of the data management.

 To overcome the limitations, we propose to use the blockchain~\cite{schrijvers2016incentive} for collecting, storing, and processing the sensing data from the SUs.~The first reason is that the blockchain is considered to be a decentralized database, i.e., a ledger~\cite{neisse2017blockchain} in which transactions, i.e., the sensing data, are recorded and processed by a number of nodes over the whole network instead of a centralized authority.~The second reason is that the blockchain enhances the security and guarantees the data integrity since the transactions must be agreed and verified by the nodes before they are recorded \cite{liu2017blockchain}.~Therefore, the blockchain can be combined with the cognitive radio to constitute a new framework called \textit{cognitive radio based blockchain network}.~The framework allows the SUs to use idle channels from the PUs to transmit their sensing data to the blockchain.~The SU transmission is likely to be ``localized'' because of inherent Device-to-Device (D2D) transmission, which is well matched to P2P connection based blockchain network.~Then, the SU's sensing data is recorded and processed in the blockchain in a decentralized but trusted manner.


However, under the dynamics of the primary channel and the uncertainty of the blockchain system, it may be challenging for each SU to make optimal decisions, i.e., transmit decision and channel selection, that maximizes the number of successful transaction transmissions.~To address the challenge, we propose to use the Deep Q-Learning (DQL) technique presented in~\cite{mnih2015human}, i.e., the combination of Deep Neural Networks (DNNs) and the Q-Learning (QL)~\cite{watkins1992q}, that enables the SU to learn the optimal policy without requiring the prior information from the network environment.~We first formulate an optimization problem for the SU that maximizes the number of successful transaction transmissions while minimizing the channel cost and transaction fee.~Then, we adopt the Double Deep Q-Network (DDQN) to implement the DQL algorithm. Simulation results show that the proposed DQL outperforms the QL in terms of the performance and learning speed.~To the best of our knowledge, this is the first work that studies the application of DQL~\cite{mnih2015human} in the cognitive radio based blockchain network.\\
The rest of this paper is organized as follows. Section~\ref{sec:system} describes the system model. Section~\ref{sec:problem} presents the problem formulation. Section~\ref{sec:dql} presents the DQL algorithm for the joint transaction transmission and channel selection in the cognitive radio based blockchain network. Section~\ref{sec:simulation} shows the performance evaluation results. Section~\ref{sec:conclusion} summarizes the paper.
\section{System Model}
\label{sec:system}
\begin{figure}[!h]
  \vspace{-0.3cm}
 \centering
\includegraphics[width=8.0cm, height = 4.2cm]{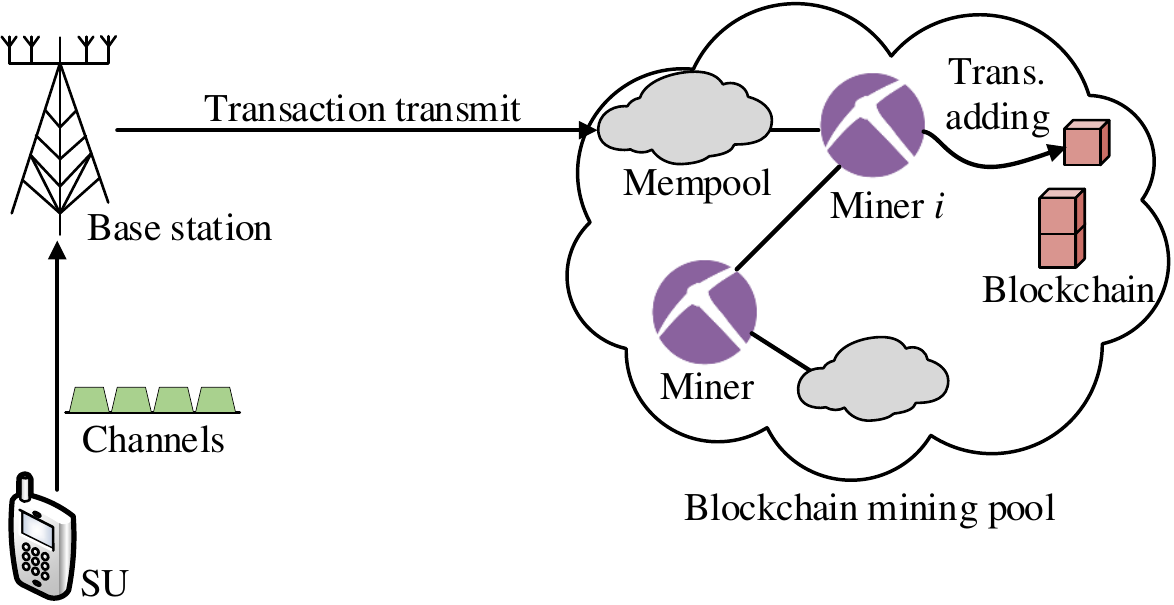}
 \vspace{-0.30cm}
 \caption{\small A cognitive radio based blockchain network.}
 \label{CRN_blockchain}
  \vspace{-0.1cm}
\end{figure}
We consider a cognitive radio based blockchain network as shown in Fig.~\ref{CRN_blockchain}.~The network consists of one SU, i.e., the IoT device, a base station, and the blockchain mining pool to support the IoT services. The base station can be regarded as a secondary receiver to establish a D2D connection with the SU. At each time slot $t$, the SU senses and selects one of $K$ channels from multiple PUs to transmit a transaction to the mining pool through a base station. Here, the channel can be good, i.e., idle, or bad, i.e., busy, due to the transmission of the corresponding PU.~The transaction contains sensing data of the SU and a transaction fee $C_T$ that the SU is willing to pay the mining pool. After being successfully verified by miners in the mining pool, the transaction is stored in a mempool of one miner, say miner $i$, as an \textit{unconfirmed transaction}. Note that the mempool is able to store $D_{\max}$ unconfirmed transactions from the SUs in the network. The miner then adds the certain number of unconfirmed transactions with high transaction fees into a new block. Thus, the probability that the transaction of the SU is successfully added to the block at the current time slot is high if the transaction is assigned with a high fee~\cite{transaction_fee}.~The blockchain system is specifically vulnerable to the double-spending attack \cite{karame2012double} in which transactions in a block can be maliciously modified. We assume that the transaction of the SU is attacked with a probability $p_a$. 
\section{Problem formulation}
\label{sec:problem}
We formulate a stochastic optimization problem for the joint transaction transmission and channel selection of the SU. The problem is defined by a tuple $<{\mathcal{S}}, {\mathcal{A}}, {\mathcal{P}}, {\mathcal{R}}>$, where ${\mathcal{S}}$, ${\mathcal{A}}$, and ${\mathcal{R}}$ are the state space, action space, and the reward function of the SU, respectively.~${\mathcal{P}}$ is the state transition probability function with $P_{s,s'}(a)$ being the probability that the current state $s \in {\mathcal{S}}$ transits to the next state $s' \in {\mathcal{S}}$ when action $a \in {\mathcal{A}}$ is executed.
\subsection{Action Space}
Let $K$ denote the number of channels that the SU can choose to transmit its transactions. Then, the action space of the SU is defined as $\mathcal{A} = \big\{ 0,1,\ldots,K  \big\}$, where $a=0$ means that the SU chooses not to transmit its transaction, and $a=k$ means that the SU chooses channel $k$ to transmit the transaction. 
\subsection{State Space}
The state space is the combination of the channel state, denoted by $\mathcal{S}^{\mathrm{c}}$, and the mempool state, denoted by $\mathcal{S}^{\mathrm{m}}$.

First, we define $\mathcal{S}^{\mathrm{c}}$. Each channel $k$ can be in one of two different states, i.e., good or bad, i.e., the channel is idle or busy because of the transmission by the PU, respectively.~The channels can be considered to be correlated, and thus all the channel states can be described as a $2^K$-state Markov chain \cite{wang2018deep} with a transition matrix $\mathbf{P}$.~At the beginning of each time slot, although the SU cannot observe the states of all the channels, it can infer the states from its past channel selections, i.e., actions, and the corresponding observations. $\mathcal{S}^{\mathrm{c}}$ is thus defined as
\begin{equation}
{\mathcal{S}}^{\mathrm{c}} = \{[a(t),w(t)], \ldots, [a(t-L+1),w(t-L+1)]\},
\end{equation}
where $w(t-l)$ is the observation of the channel selection at time slot $t-l$. $w(t-l)=1$ if the channel is good, and $w(t-l)=0$ if the channel is bad. $L$ is the number of observations. 

Second, we define mempool state $\mathcal{S}^{\mathrm{m}}$.~$\mathcal{S}^{\mathrm{m}}$ refers to the current number of transactions and the corresponding transaction fees in the mempool.~$\mathcal{S}^{\mathrm{m}}$ is defined as ${\mathcal{S}}^{\mathrm{m}} = \{(m_1, \Delta C_1), \ldots, (m_M,\Delta C_M)\}$, where $\Delta C_i$ represents transaction fee range $i$, and $m_i$ is the current number of transactions that have transaction fees within the range of $\Delta C_i$. The transition of the mempool state from time slot $t$ to $t+1$ depends on (i) the number of transactions arriving in the mempool, (ii) the corresponding transaction fees, and (iii) the number of transactions that the miner adds a new block at time slot $t$.~Here, we assume that the number of transactions arriving in the mempool and the transaction fee follow the uniform distributions $U[T^{\min}, T^{\max}]$ and $U[C^{\min}_T, C^{\max}_T]$, respectively.~The number of transactions added to the new block also follows the uniform distribution $U[T^{\min}_{\text{add}},T^{\max}_{\text{add}}]$. 

The state space of the SU at time slot $t$ is thus defined as ${\mathcal{S}}	=	{\mathcal{S}}^{\mathrm{c}} \times {\mathcal{S}}^{\mathrm{m}}$, where $\times$ is the Cartesian product. 
\subsection{Reward Function}
The reward function $\mathcal{R}$ of the SU is composed of three components, i.e., the positive utility $R_{\text{success}}$, the channel access cost $C_c$ , and the transaction fee $C_T$.~The SU receives the utility of $R_{\text{success}} > 0$  if the transaction transmission is successful and the utility of $R_{\text{success}} = 0$ otherwise. The transaction transmission is considered to be ``successful'' if it is added to the new block at the current time slot and is not attacked. Here, we introduce the double-spending attack in which the transaction is attacked with a probability $p_a$ given by~\cite{rosenfeld2014analysis}:
\begin{equation}
p_a = \left\{\begin{matrix}
1-\sum_{m=0}^{n} \begin{pmatrix}
m+n-1 \\ m
\end{pmatrix} 
\left ( p^n q^m-p^m q^n \right ) & \text{if  } q<p,\\
1 & \text{if  } q \geq p,
\end{matrix}\right.
\notag
\end{equation}
where $n$ and $m$ are respectively the numbers of blocks that are found by the honest network and the attacker, $p$ and $q$, $p+q=1$, are the probabilities that a block is found by the honest network and the attacker, respectively.

The objective is to maximize the number of successful transaction transmissions and minimize the channel cost and the transaction fee. Thus, the reward function of the SU is defined as $\mathcal{R}(s,a)= R_{\text{success}} -C_c -  C_T$.

To obtain the mapping from a state $s \in {\mathcal{S}}$ to an action $a \in {\mathcal{A}}$ such that the long-term accumulated reward is maximized, the QL algorithm can be used.~The algorithm finds the optimal policy defined as $\pi^* : {\mathcal{S}} \rightarrow {\mathcal{A}}$ by estimating Q-values of state-action pairs, i.e., $Q(s,a)$.~$Q(s,a)$ is the expected discounted sum of future rewards obtained by taking an action $a$ at state $s$ following the optimal policy.~The Q-values are updated based on the experience of the SU as follows: 
\begin{align}
\label{Q_value_update}
Q^{\mathrm{new}}(s,a) = & (1-\lambda) Q(s,a) \\
& + \lambda \left( r(s,a) + \gamma \max_{a' \in {\mathcal{A}} } Q( s', a') \right),\notag
\end{align}
where $\lambda$ is the learning rate, and $\gamma$ is the discount factor.

After the values $Q(s,a)$ are learned, the SU can determine its optimal action from any state to maximize the long-term accumulated reward.~However, the QL suffers from large state and action spaces of the network. Thus, we propose to use a DQL to find the optimal policy for the SU. 

\section{Deep Q-Learning Algorithm}
\label{sec:dql}

DQL uses a DNN instead of the look-up table to represent all the states and actions of the SU. The input of the DNN is one of the states of the SU, and the output includes Q-values of all possible actions. To enable the SU to map its current state to an optimal action, the DNN needs to be trained. Training the DNN is to update its weights $\boldsymbol{\theta}$ by using experiences $e=< s,a,r,s'>$ of the SU to minimize a loss function. Here, the SU can execute action $a$ using the $\epsilon$-greedy policy to balance its exploration and exploitation. The loss function at the current iteration is given by $L= \mathbb{E}\left[ (y(t)-Q(s,a;\boldsymbol{\theta}))^2\right]$, where $y$ is the target value. Typically, $y$ is defined as $y= r+ \gamma \max_{a' \in {\mathcal{A}} }  Q( s', a';\boldsymbol{\theta^{\text{old}}})$, where $\boldsymbol{\theta^{\text{old}}}$ are the weights of the DNN at the last iteration. However, such definition results in over-optimistic value estimates since the $\max$ operator in $y$ uses the same Q-values both to select and to evaluate an action. To decouple the action selection from the action evaluation, we propose to use the DDQN~\cite{van2016deep} which is composed of one online DNN with weights $\boldsymbol{\theta^{\text{online}}}$ and one target DNN with weights $\boldsymbol{\theta^{\text{target}}}$.~The online DNN updates its weights $\boldsymbol{\theta^{\text{online}}}$ at each iteration.~The target DNN resets its weights $\boldsymbol{\theta^{\text{target}}}$ to $\boldsymbol{\theta^{\text{online}}}$ in every $N^{\text{target}}$ iterations and keeps weights $\boldsymbol{\theta^{\text{target}}}$ fixed at other iterations.~The online DNN updates its weights $\boldsymbol{\theta}$ to minimize the loss function defined as
\begin{equation}
L^{\text{DDQN}}= \mathbb{E}\left[ (y^{\text{DDQN}}-Q(s,a;\boldsymbol{\theta^{\text{online}}}))^2\right], 
\label{DDQN_loss}
\end{equation}
where the target value $y^{\text{DDQN}}$ is defined as
\begin{equation}
y^{\text{DDQN}}=r + \gamma Q\Big{(} s', \arg\max_{a' \in \mathcal{A}} Q_i(s',a';\boldsymbol{\theta^{\text{online}}});\boldsymbol{\theta^{\text{target}}}\Big{)}.
\label{DQN_y_value_DDQN}
\end{equation}

(\ref{DQN_y_value_DDQN}) shows that the selection of an action is due to the current weights, i.e., $\boldsymbol{\theta^{\text{online}}}$, while the weights $\boldsymbol{\theta^{\text{target}}}$ of the target DNN are used to evaluate fairly the value of the action.
\begin{algorithm}
\footnotesize 	
\caption{\small DQL algorithm~\cite{mnih2015human}.}\label{DDQN_Algorithm}
\hspace*{\algorithmicindent} \textbf{Input:} $\mathcal{A}$; $N^{\text{target}}$; $N_b$; $\mathcal{M}$\\
\hspace*{\algorithmicindent} \textbf{Output:} Optimal policy $\pi^*$
\begin{algorithmic}[1]
\State \textbf{Initialize:} $\boldsymbol{\theta^{\text{online}}}$; $\boldsymbol{\theta}^{\text{target}} $
\For{\texttt{episode $i$ = $\{1,\ldots,N \}$}}
\For{\texttt{iteration $t= \{1,\ldots,T \}$}}

\State Execute action $a$ according to $\epsilon-greedy$ policy
\State Receive reward $r_t$
\State Store experience $(s, a, r_t, {s}')$ in $\mathcal{M}$
\State Sample $N_b$ experiences $(s, a, r_j, {s}')$ from $\mathcal{M}$
\If{ if an episode terminates at iteration $j + 1$}
    \State Set $y_j^{\text{DDQN}}=r_j$
  \Else
  \State Determine $a^{\text{max}}=\arg \max_{a' \in \mathcal{A}} Q({s}', {a}'; \boldsymbol{\theta}^{\text{online}})$
    \State Set $y_j^{\text{DDQN}}=r_j+ \gamma Q\left ({s}', a^{\text{max}};\boldsymbol{\theta}^{\text{target}} \right)$
  \EndIf
\State Perform a gradient descent step on $(y_{j}^{\text{DDQN}}-Q(s,a;\boldsymbol{\theta}^{\text{online}}))^{2}$ to update $\boldsymbol{\theta}^{\text{online}}$

\State Reset $\boldsymbol{\theta}^{\text{target}} = \boldsymbol{\theta}^{\text{online}}$ in every $N^{\text{target}}$ iterations

\EndFor 
\EndFor
\end{algorithmic}
\end{algorithm}

Algorithm~\ref{DDQN_Algorithm} shows the DQL algorithm which uses the DDQN to find the optimal policy for the SU. Accordingly, based on the experience $e$, the online DNN and target DNN compute the optimal value $Q(s',a';\boldsymbol{\theta^{\text{online}}})$. Then, the target value $y^{\text{DDQN}}$ and the loss function $L^{\text{DDQN}}$ is calculated according to (\ref{DQN_y_value_DDQN}) and (\ref{DDQN_loss}), respectively. The value of $L^{\text{DDQN}}$ is used to update weights $\boldsymbol{\theta}$ of the online DNN. To ensure the stability of the learning, the experience replay memory $\mathcal{M}$ is used to store experience $e$, and then a mini-batch of $N_b$ experiences are taken at each iteration to train the DNNs. 
\section{Performance Evaluation}
\label{sec:simulation}
\begin{figure}[h]
  \vspace{-0.1cm}
 \centering
\includegraphics[width=6.3cm, height = 4.7cm]{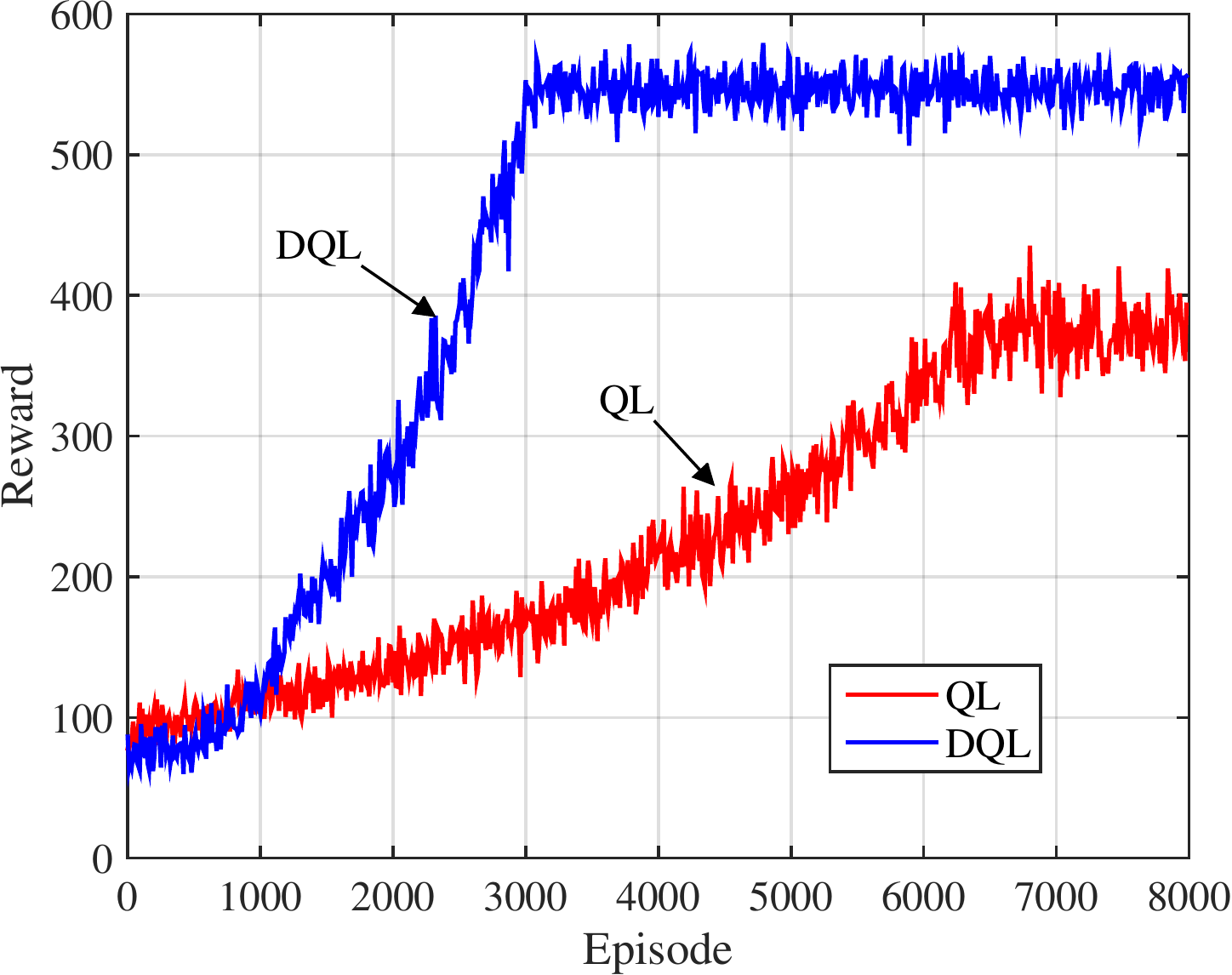}
  \vspace{-0.3cm}
 \caption{\small DQL scheme and QL scheme comparison.}
  \label{DQN_versus_QL}
  \vspace{-0.1cm}
\end{figure}

\begin{figure}[h]
  \vspace{-0.1cm}
 \centering
\includegraphics[width=6.3cm, height = 4.7cm]{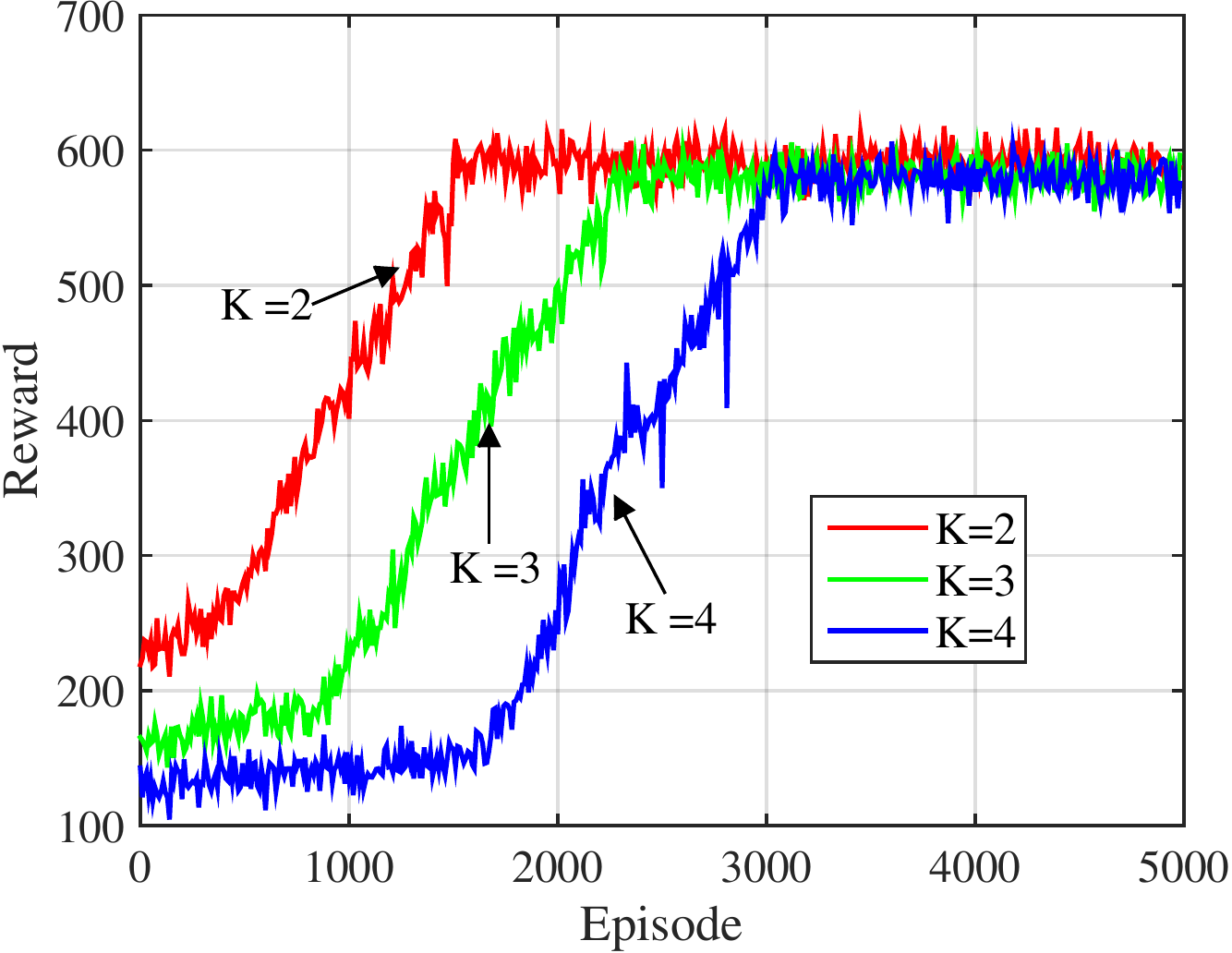}
 \vspace{-0.35cm}
 \caption{\small Reward of DQL as the number of channels $K$ is varied.}
  \label{number_channel_changing}
   \vspace{-0.1cm}
\end{figure}

\begin{figure}[!h]
\vspace{-0.05cm}
 \centering
\includegraphics[width=6.3cm, height = 4.7cm]{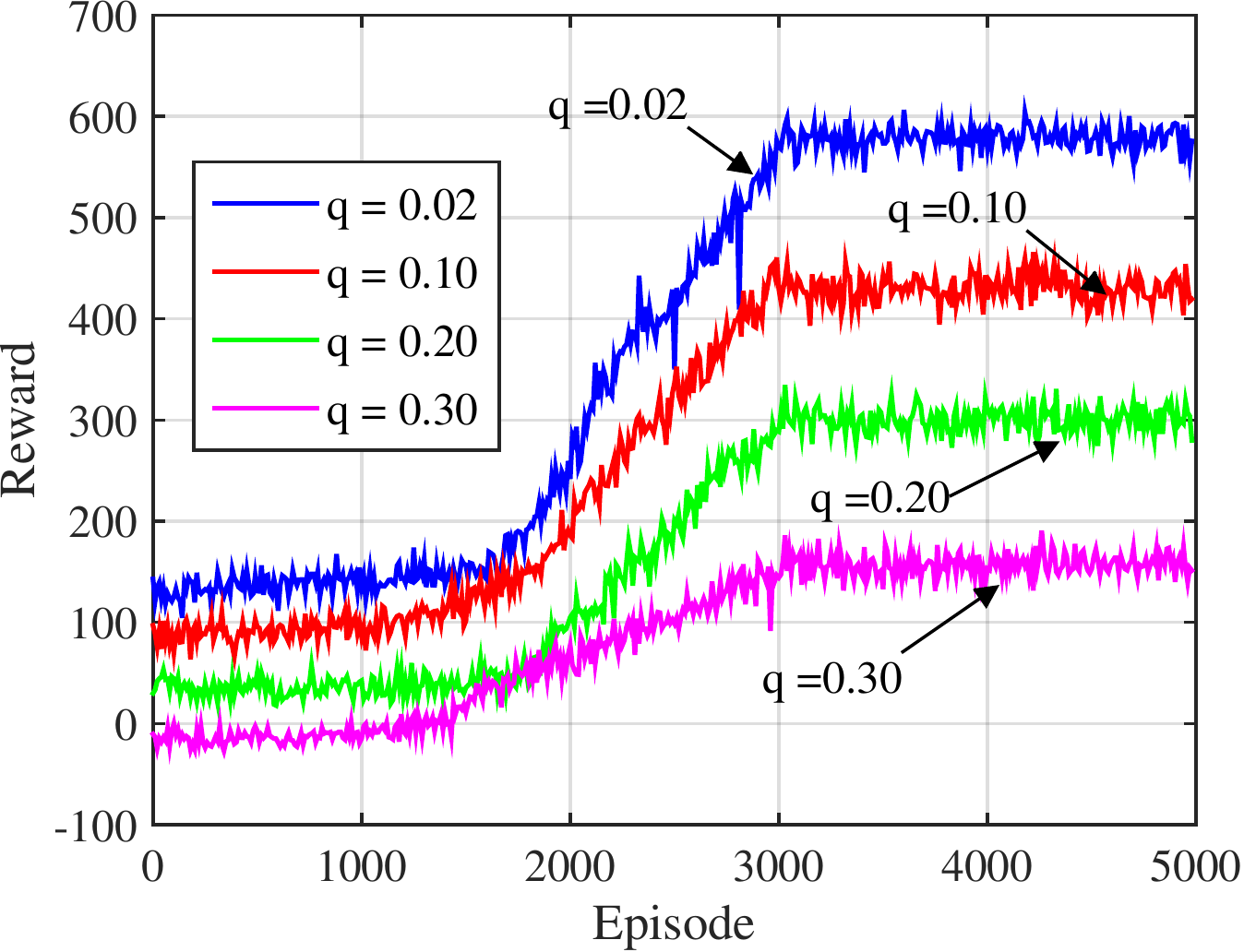}
 \vspace{-0.2cm}
 \caption{\small Reward of DQL as the probability $q$ is varied.}
 \label{attack_changing}
  \vspace{-0.1cm}
\end{figure}

\begin{figure}[h]
  \vspace{-0.1cm}
 \centering
 \includegraphics[width=6.3cm, height = 4.5cm]{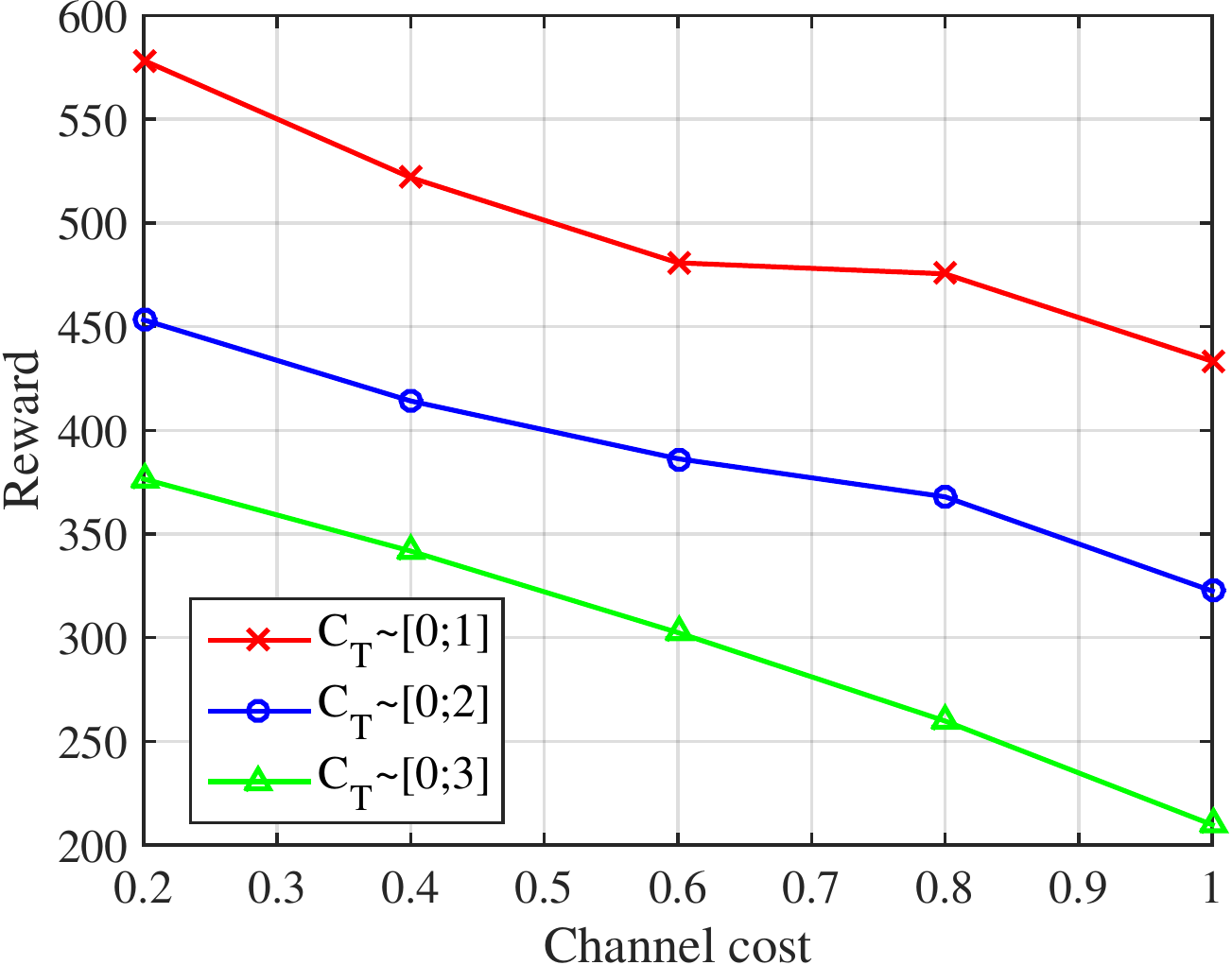}
 \vspace{-0.3cm}
 \caption{\small Reward of DQL versus the channel access cost $C_c$.}
  \label{ChannelCost_and_Transaction_fee_changing}
   \vspace{-0.5cm}
\end{figure}

In this section, we present experimental results to evaluate the performance of the proposed DQL algorithm. For comparison, the QL algorithm \cite{watkins1992q} is used as a baseline scheme. Major simulation parameters are listed in Table~\ref{table:parameters_CRN}.~The simulation results for the performance comparison between the proposed DQL scheme and the QL scheme are shown in Figs.~\ref{DQN_versus_QL},~\ref{number_channel_changing},~\ref{attack_changing}, and~\ref{ChannelCost_and_Transaction_fee_changing} depending on the varied parameters. 

\begin{table}[h]
  \vspace{-0.3cm}
\small
\caption{\small Simulation parameters}
\label{table:parameters_CRN}
\footnotesize 	
\centering
\begin{tabular}{lc}
\hline\hline
{\em Parameters} & {\em Value} \\ [0.5ex]
\hline
Number of channels             ($K$)    & $4$             \\ 
Probability of switching good channel ($p_c$)& $0.9$             \\
Maximum number of transactions in the mempool ($D_{\max}$)       & $50$    \\ 
Channel cost $(C_c)$							& $0.2$ \\
Transaction fee	 $(C_T)$						&$\sim U[0;1]$ \\
Probability that a block is found by the attacker $(q)$	            & $0.02$   \\ 
Discount rate ($\gamma$) & 0.9 \\
$\epsilon$-greedy & 0.9 $\rightarrow$ 0 \\
\hline
\end{tabular}
\label{table:parameters}
\vspace{-0.3cm}
\end{table}

Fig.~\ref{DQN_versus_QL} illustrates the rewards obtained by the DQL and QL schemes.~To enable the QL scheme to be run in our computation environment, we reduce the state space by setting the maximum number of transactions in the mempool to be $10$.~As seen, the DQL scheme converges to the reward much higher than that of the QL scheme. Specifically, the rewards obtained by the DQL and QL schemes are $550$ and $390$, respectively. Moreover, the convergence speed of the DQL scheme is faster than that of the QL scheme.~The DQL scheme converges at around $3000$ episodes while the QL scheme converges at $7000$ episodes.

The convergence speed of the DQL scheme is likely maintained as the maximum number of transactions in the mempool increases to $50$ as shown in Fig.~\ref{number_channel_changing}. In this case, the state space becomes too large for the QL scheme to converge in the reasonable time, and hence it is not shown in the figure. This confirms the scalability of the DQL. Note that as the number of channels $K$ is varied, the state space changes, and the DQL scheme has different convergence speeds.~However, the DQL scheme always reaches to the same reward because it already learns the optimal policy to obtain the maximum reward.   

The reward obtained by the DQL scheme decreases as the probability that a block is found by the attacker $q$ increases as shown in Fig.~\ref{attack_changing}.~The reason is that as $q$ increases, the number of successfully transmitted transactions decreases. Similarly, the reward that the SU receives decreases as the transaction fee $C_T$ and the channel cost $C_c$ increase as illustrated in Fig.~\ref{ChannelCost_and_Transaction_fee_changing}.  

\section{Conclusions}
\label{sec:conclusion}

In this paper, we have presented the DQL algorithm for the joint transaction transmission and channel selection problem in the cognitive radio based blockchain network. Specifically, we have developed a DQL algorithm using DDQN to solve the problem.~The
simulation results show that the proposed DQL scheme outperforms the QL scheme in terms of reward and learning speed. This implies that the DQL enables the SU to achieve the higher number of successful transaction transmissions while paying lower cost. 

\vfill\pagebreak

\bibliographystyle{IEEEtran}
\bibliography{strings,refs}

\end{document}